\documentclass[aip,jmp]{revtex4-1}
\usepackage{graphicx}
\usepackage{amsmath}
\usepackage{amsfonts}
\usepackage{amssymb}
\usepackage{MnSymbol}

\newcommand{\be}{\begin{equation}}
\newcommand{\ee}{\end{equation}}
\newcommand{\tr}{{\rm Tr}}
\newcommand{\Wg}{{\rm Wg}}

\begin{document}

\title{Statistics of time delay and scattering correlation functions in chaotic systems II. Semiclassical Approximation}
\author{Marcel Novaes}
\affiliation{Instituto de F\'isica, Universidade Federal de Uberl\^andia\\ Av. Jo\~ao
Naves de \'Avila 2121, Uberl\^andia, MG, 38408-100, Brazil}
\begin{abstract}
We consider $S$-matrix correlation functions for a chaotic cavity having $M$ open
channels, in the absence of time-reversal invariance. Relying on a semiclassical approximation,
we compute the average over $E$ of the quantities ${\rm Tr}[S^\dag(E-\epsilon)S(E+\epsilon)]^n$, for general
positive integer $n$. Our result is an infinite series in $\epsilon$, whose coefficients are
rational functions of $M$. From this we extract moments of the time delay matrix $Q=-i\hbar S^\dag dS/dE$,
and check that the first 8 of them agree with the random matrix theory prediction from our previous paper [M. Novaes, submitted].
\end{abstract}
\maketitle

\section{Introduction}

Quantum scattering processes at energy $E$ can be described by the scattering matrix
$S(E)$, which transforms incoming wavefunctions into outgoing wavefunctions. This matrix
is necessarily unitary, in order to enforce conservation of probability and,
consequently, conservation of charge. We consider a scattering region (`cavity') inside
of which the classical dynamics is strongly chaotic, connected to the outside world by
small, perfectly transparent, openings. We assume $M$ open channels, so that $S$ is
$M$-dimensional. We also assume there is a well defined classical decay rate $\Gamma$,
such that the total probability of a particle to be found inside the cavity decays
exponentially in time as $ e^{-\Gamma t}$. The quantity $\tau_D=1/\Gamma$ is called the
classical `dwell time'.

The study of time delay requires the energy dependence of the $S$ matrix, as the
Wigner-Smith time delay matrix \cite{wigner,smith} is defined as $Q=-i\hbar S^\dag
dS/dE$. Properties of $Q$ may also be derived starting from correlation functions
\be\label{correlation} C_n(\epsilon,M)=\frac{1}{M}\left\langle{\rm Tr}\left[
S^\dag\left(E-\frac{\epsilon \hbar}{2\tau_D}\right)S\left(E+\frac{\epsilon
\hbar}{2\tau_D}\right)\right]^n\right\rangle,\ee where $\langle\cdot\rangle$ denotes an
average over $E$. For example, the average value of time delay moments
$\mathcal{M}_m=\frac{1}{M}\tr(Q^m)$ can be obtained as \cite{berko1} \be\label{MfromC} \langle
\mathcal{M}_m\rangle=\frac{\tau_D^m}{i^mm!}\left[\frac{d^m}{d\epsilon^m}\sum_{n=1}^m
(-1)^{m-n}{m \choose n}C_n(\epsilon)\right]_{\epsilon=0}. \ee

In the semiclassical regime (when $\hbar\to 0$ and the electron wavelength is much
smaller than the cavity size), we may use the semiclassical approximation, in which
elements of the $S$ matrix are written as sums over classical scattering trajectories
\cite{jalabert}. Calculation of energy-averaged transport statistics (like condutance,
shot-noise, etc) then require so-called action correlations, sets of trajectories having
the same total action, leading to constructive interference. Using only identical
trajectories and ergodicity arguments \cite{berry,cvitanovic,hannay} one can recover some
semiclassical large-$M$ asymptotics. Quantum corrections, important at finite $M$, can be
related to non-identical trajectories having close encounters \cite{sieber}, and may be
obtained systematically \cite{muller,greg,novaes1}.

The semiclassical approach can also be used to study time delay. Interestingly, in this
case one can use the periodic orbits \cite{balian} that live in the fractal chaotic
saddle of the system \cite{altmann} (sometimes called `the repeller'). This approach was
followed in \cite{eckhardt,vallejos,KS1}. It is actually equivalent \cite{KS2} to the one
based on scattering trajectories \cite{caio3}, which Berkolaiko and Kuipers used to treat
(\ref{correlation}) semiclassically, initially in the large-$M$ limit \cite{berko1} and
later up to the first finite-$M$ corrections. \cite{berko2} Another semiclassical
approach to time delay, that avoids correlation functions, has recently been introduced.
\cite{new}

We hereby advance the semiclassical approach by deriving a formula for correlation
functions $C_n(\epsilon,M)$ which is a Taylor series in $\epsilon$, with coefficients
that are rational functions of $M$. These coefficients are expressed as finite sums
involving characters of the symmetric group and Stirling numbers. Our method is an
extension of a recently introduced semiclassical matrix model for transport statistics.
\cite{novaes2}

Statistical properties of $Q$ can also be calculated using random matrix theory (RMT).\cite{sigma1,sigma2,sigma3,sigma4}
In our previous paper,\cite{previous} we obtained the average
value of general polynomial functions of $Q$. Equivalence between the semiclassical and
RMT approach has long been conjectured, and was previously known to be true to leading
orders in $1/M$. The structure of our formula for $C_n(\epsilon,M)$ suggests that this
equivalence holds exactly in $M$ for all polynomial functions of $Q$. We are able to
verify this in many cases, but come short of showing it in full generality.

We remark that the semiclassical approximation provides the energy-dependent correlation
functions, which have more information than the energy-independent RMT statistics
obtained in \cite{previous}. For instance, correlation functions are required in order to
develop a semiclassical treatment of Andreev systems. \cite{andreev1,andreev2} Also, the
semiclassical approximation is in principle able to go beyond RMT by including Ehrenfest
time effects. \cite{ehren1,ehren2,ehren3,ehren4} These developments are outside
the scope of the present work, but we hope they will attract attention in the future.

A last remark about our semiclassical calculation. It is based on an integral over
$N$-dimensional complex matrices, and requires that we take the limit $N\to 0$. This
limit is needed to enforce that our semiclassical expansions do not contain periodic
orbits. It is easily taken in the perturbative framework (see Section III.B), i.e. order by
order in $1/M$. However, we cannot rigorously justify it for the exact calculation.
The same issue exists for transport statistics.
\cite{novaes2} We
believe the nature of this limit is an interesting open problem that deserves further
study.

This paper is organized as follows. In the next Section we present and discuss our
results. In Section 4 we present our calculations. They rely on some well known facts
about symmetric functions and the permutation group, which we have reviewed in our
previous paper. \cite{previous}

\section{Results}

We develop a new formulation for the semiclassical approach to time delay, based on our
previous work on transport statistics \cite{novaes2}. This requires a matrix integral
which is designed to have the correct diagrammatic expansion, so that it mimics the
semiclassical approximation to the correlation functions $C_n(\epsilon,M)$.

Solving exactly that matrix integral, we arrive at a formula for $C_n(\epsilon,M)$ in the
form of a Taylor series in $\epsilon$. Let $\chi_\lambda(\mu)$ be the characters of the
irreducible representations of the permutation group and $d_\lambda=\chi_\lambda(1^n)$ be
the dimension of such a representation (we have reviewed these concepts in our previous
paper \cite{previous}). Our formula is \be C_n(\epsilon,M)=\frac{1}{Mn!}\sum_{m=0}^\infty
\frac{(Mi\epsilon)^m}{m!}\sum_{\lambda\vdash n}\sum_{\mu\vdash m}d_\lambda
d_\mu\chi_\lambda(n)\frac{[M]^\lambda}{[M]_\mu}F_{\lambda,\mu},\ee where
$F_{\lambda,\mu}$ is some complicated function for which we have an explicit form (see
Section 4.4.1), and \be\label{factorials}
[M]^\lambda=\prod_{i=1}^{\ell(\lambda)}[M-i+1]^{\lambda_i},\quad
[M]_\lambda=\prod_{i=1}^{\ell(\lambda)}[M+i-1]_{\lambda_i}\ee are generalizations of the
rising and falling factorials.

For the simplest correlation function, explicit calculations suggest that the following
expression holds: \be\label{C1} C_1(\epsilon)=\sum_{n=1}^\infty
\frac{(Mi\epsilon)^n}{n}\sum_{k=0}^{n-1}\frac{1}{[M+k]_n}.\ee We can establish that the leading
order in $\epsilon$ is given, for any $n$, by $C_n(\epsilon,M)=1+ni\epsilon+O(\epsilon^2)$.

To leading orders in $1/M$ we find, for example, that \be
C_1=\frac{1}{1-i\epsilon}-\frac{\epsilon^2}{M^2(1-i\epsilon)^5}-
\frac{\epsilon^2(1+12i\epsilon-8\epsilon^2)}{M^4(1-i\epsilon)^9}+O(1/M^6),\ee which is
indeed in agreement with the first $3$ orders as computed from (\ref{C1}). For
the second correlation function we do not have a simple formula, but we can show that
\be\label{C2} C_2=\frac{(1-2i\epsilon-2\epsilon^2)}{(1-i\epsilon)^4}-
\frac{\epsilon^2(4+8i\epsilon-7\epsilon^2-2i\epsilon^3)}{M^2(1-i\epsilon)^8}+O(1/M^4).\ee
This generalizes some results that appear in the Appendix of \cite{berko1}.

The average value of moments $\mathcal{M}_m$ have been computed semiclassically up to
the first few orders in $1/M$. \cite{berko2} Using our new exact semiclassical expression
for $C_n(\epsilon,M)$, we could compute them in closed form as rational functions of $M$
up to $m=8$, and check that they agree with RMT predictions from \cite{previous}. Unfortunately, we could
not establish this agreement in general, because of the complicated nature of the
function $F_{\lambda,\mu}$.

\section{Semiclassical Approach to Correlation Functions}

\subsection{Semiclassical Approximation}

In the semiclassical limit $\hbar\to 0$, $M\to \infty$, the element $S_{oi}$ of the $S$
matrix may be approximated by a sum over trajectories $\gamma$ starting at channel $i$
and ending at channel $o$: \cite{jalabert} \be
S_{oi}=\frac{1}{\sqrt{T_H}}\sum_{\gamma:i\to o}A_\gamma e^{i\mathcal{S}_\gamma/\hbar}.\ee
The phase $\mathcal{S}_\gamma$ is the action of $\gamma$, while $A_\gamma$ is related to
its stability. The prefactor contains the so-called Heisenberg time, $T_H=M\tau_D$.

Consider the correlation function $C_n(\epsilon,M)=\frac{1}{M}\left\langle{\rm Tr}\left[
S^\dag\left(E-\frac{\epsilon \hbar}{2\tau_D}\right)S\left(E+\frac{\epsilon
\hbar}{2\tau_D}\right)\right]^n\right\rangle$. Expanding the trace, we find a multiple
sum over trajectories, \be\label{multiple}
C_n(\epsilon,M)=\frac{1}{MT_H^n}\prod_{k=1}^n\sum_{i_k,o_k}\sum_{\gamma_k,\sigma_k}A_\gamma
A^*_\sigma
e^{i(\mathcal{S}_\gamma-\mathcal{S}_\sigma)/\hbar}e^{\frac{i\epsilon}{2\tau_D}(T_\gamma+T_\sigma)},\ee
such that $\gamma_k$ goes from $i_k$ to $o_k$, while $\sigma_k$ goes from $i_k$ to
$o_{k+1}$. The channels labels are all being summed from $1$ to $M$.

In (\ref{multiple}) we have used \be \mathcal{S}_\gamma(E+\frac{\epsilon
\hbar}{2\tau_D})\approx \mathcal{S}_\gamma(E)+\frac{\epsilon \hbar}{2\tau_D}T_\gamma,\ee
where $T_\gamma$ is the total duration of $\gamma$. The quantity $A_\gamma=\prod_k
A_{\gamma_k}$ is a collective stability, while $\mathcal{S}_\gamma=\sum_k
\mathcal{S}_{\gamma_k}$ and $T_\gamma=\sum_k T_{\gamma_k}$ are the collective action and
duration of the $\gamma$ trajectories, and analogously for $\sigma$.

The result of the sum in (\ref{multiple}) is, for a chaotic system, a strongly
fluctuating function of the energy. The average over $E$, under the stationary phase
approximation, requires $\gamma$ and $\sigma$ to have almost the same collective action.
In the past years \cite{sieber}, it has been established that these action correlations
arise when each $\sigma$ follows closely a certain $\gamma$ for a period of time, and
some of them exchange partners at so-called encounters. A $q$-encounter is a region where
$q$ pieces of trajectories run nearly parallel and $q$ partners are exchanged. This
theory has been presented in detail before. \cite{muller,R2c} We consider only systems
not invariant under time-reversal, so $\sigma$ trajectories never run in the opposite
sense with respect to $\gamma$ trajectories.

\begin{figure}
\includegraphics[scale=0.7,clip]{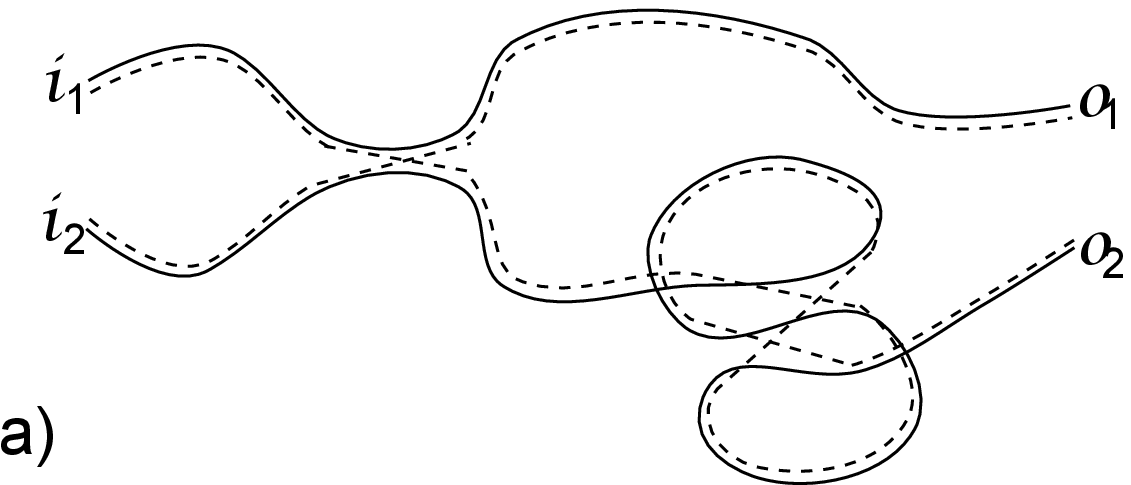}\qquad\includegraphics[scale=0.7,clip]{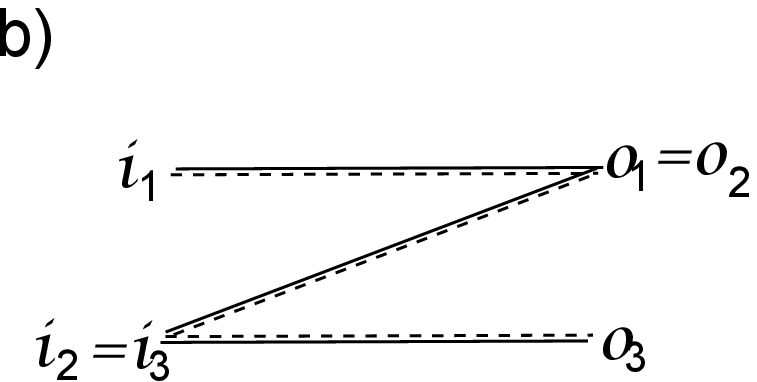}
\caption{a) Correlated trajectories contributing to $C_2(\epsilon,M)$. Solid lines are
$\gamma_1$ (going from $i_1$ to $o_1$) and $\gamma_2$ (going from $i_2$ to $o_2$), dashed
lines are $\sigma_1$ (going from $i_1$ to $o_2$) and $\sigma_2$ (going from $i_2$ to
$o_1$). In this situation we have one $2$-encounter and one $3$-encounter (the encounters
are greatly magnified). b) Correlated trajectories contributing to $C_3(\epsilon,M)$, in
a case with coinciding channels. In both figures the chaotic nature of the trajectories
is not shown.}
\end{figure}

For example, we show in Figure 1a a situation contributing to the second correlation
function, $C_2(\epsilon,M)$. Trajectory $\gamma_1$ starts in channel $i_1$ and ends in
channel $o_1$, while $\gamma_2$ starts in channel $i_2$ and ends in channel $o_2$. On the
other hand, $\sigma_1$ and $\sigma_2$ are initially almost identical to $\gamma_1$ and
$\gamma_2$, respectively, but they exchange partners in a $2$-encounter. Later,
$\gamma_2$ has a $3$-encounter with itself, inside of which the pieces of $\sigma_1$ are
connected differently. We also show in Figure 1b a situation contributing to
$C_3(\epsilon,M)$ which has no encounters, but has coinciding channels. There are two
major simplifications done here for visual clarity: 1) The encounters are greatly
magnified, to show their internal structure; 2) The actual trajectories are extremely
convoluted and chaotic. Many other examples of correlated trajectories can be found in
previous work. \cite{sieber,muller,greg,novaes1,berko1,berko2,novaes2}

Correlated sets of trajectories contributing to the semiclassical calculation of
correlation functions can be depicted in the form of ribbon graphs, as suggested in
\cite{GregJack1,GregJack2}. The $q$-encounters become vertices of valence $2q$. Channels
also become vertices, but their valence depends on whether there are coinciding channels
or not. The pieces of trajectories connecting vertices become fat edges, or ribbons. Each
ribbon is bordered by one $\gamma$ and one $\sigma$, and these trajectories traverse the
encounter vertices in a well defined rotation sense: a trajectory arriving from one
ribbon departs via the adjacent ribbon (graphs endowed with a cyclic order around
vertices are also called maps). We show in Figure 2 the ribbon graphs corresponding to
the trajectories shown in Figure 1.

\begin{figure}
\includegraphics[scale=0.75,clip]{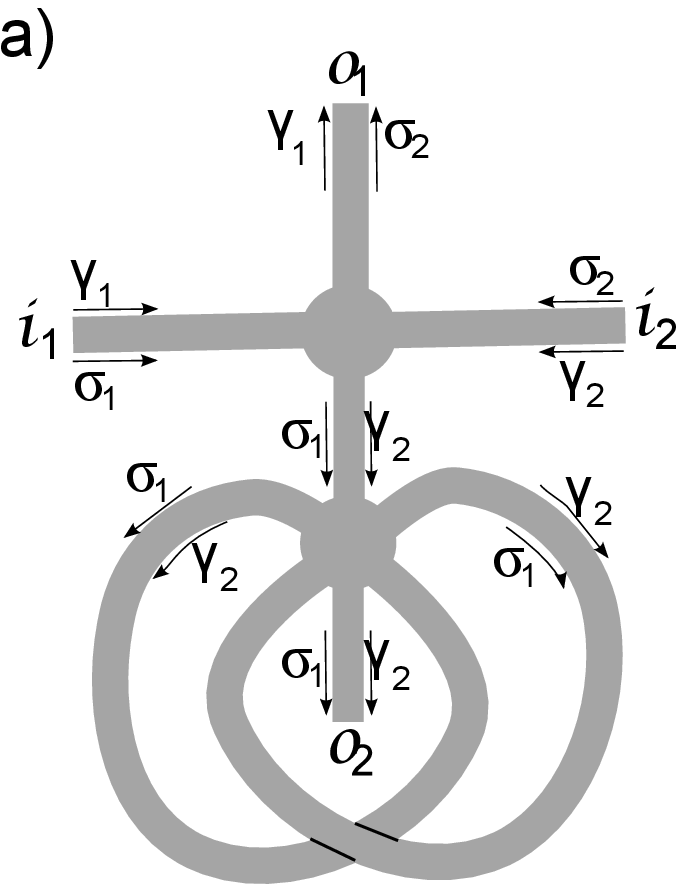}\qquad\includegraphics[scale=0.75,clip]{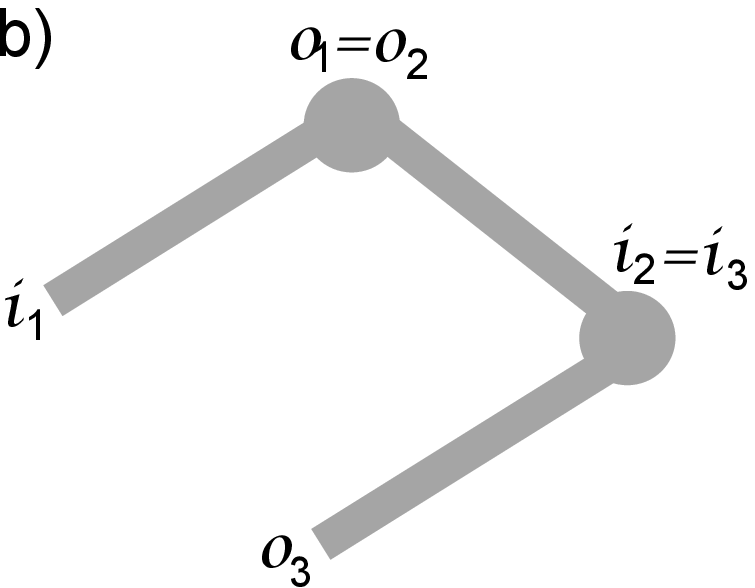}
\caption{The ribbon graphs corresponding to Figure 1. Each ribbon is bordered by one
$\gamma$ and one $\sigma$. Ribbons only meet at vertices, and $q$-encounters become
vertices of valence $2q$.}
\end{figure}

Following previous work on transport and on closed systems, Kuipers and Sieber obtained
some diagrammatic rules \cite{KS2}, that determine how much a given graph contributes to
the correlation function. The contribution of a graph factorizes into the contributions
of individual vertices and edges: an encounter vertex of valence $2q$ gives rise to
$-M(1-iq\epsilon)$; channels of any valence give rise to $M$; each ribbon gives rise to
$[M(1-i\epsilon)]^{-1}$. These rules were then used in several works dealing with time
delay statistics \cite{berko1,berko2,andreev1,andreev2}.

Notice that there are no periodic orbits in a ribbon graph that arises from the
semiclassical expansion of time delay. This means that we may start from $i_n$ and follow
$\sigma_1$ up to $o_1$, then follow $\gamma_1$ in reverse back to $i_1$, then $\sigma_2$
to $o_2$, then $\gamma_2$ in reverse back to $i_2$, and so on, and traverse every border
of every ribbon exactly once. This means that the graph has a single face.

The contribution of a graph will be proportional to $M^{V-E-1}$, where $V$ is the total
number of vertices (including channels) and $E$ is the total number of edges. The Euler
characteristic of a ribbon graph is $V-E+F$, where $F$ is the number of faces ($F=1$ in
our case). The Euler characteristic is also equal to $2-2g$, where $g$ is called the
genus. Therefore, the $1/M$ expansion coming from semiclassical diagrammatics is actually
what is called a genus expansion: the contribution of a graph is proportional to
$1/M^{2g}$. Graphs with $g=0$ are called planar (they can be drawn on the plane so that
the ribbons never cross each other), and they give the leading order contribution.

The graph in Figure 2a, for example, contributes \be
\frac{(1-2i\epsilon)(1-3i\epsilon)}{M^2(1-i\epsilon)^7}\ee to $C_2(\epsilon,M)$. Notice
that it is not a planar graph, since there is a crossing between two of the ribbons. This
particular graph actually has $g=1$ (this means it may be drawn on a torus without any
crossings). The graph in Figure 2b, on the other hand, is planar and contributes
$(1-i\epsilon)^{-3}$ to $C_3(\epsilon,M)$.

\subsection{Gaussian integrals and Wick diagrammatics}

We shall introduce a certain Gaussian matrix integral and formulate it diagrammatically,
using Wick's rule. This procedure has been discussed in detail for hermitian matrices.
\cite{difrancesco,bouttier} The only difference compared to the present work is that we
integrate over non-hermitian matrices. Our diagrams are then interpreted as providing the
semiclassical formulation of the time delay problem. The same approach has been used to
treat transport statistics. \cite{novaes2}

Let $Z$ denote a general complex matrix of dimension $N$, and define \be\label{ave}
\llangle f(Z,Z^\dag)\rrangle\equiv \frac{1}{\mathcal{Z}}\int dZ
e^{-\Omega\tr(ZZ^\dag)}f(Z,Z^\dag), \ee where the normalization constant is \be
\mathcal{Z}=\int dZ e^{-\Omega\tr(ZZ^\dag)}.\ee We see (\ref{ave}) as an average value,
but we use the symbol $\llangle\cdot\rrangle$ to differentiate it from the true physical
energy-average we considered in previous sections. For example, since the elements are
actually independent, it is clear that \be\label{cov} \llangle
Z_{mj}Z^\dag_{qr}\rrangle=\frac{\delta_{mr}\delta_{jq}}{\Omega}.\ee

Integrals over a product of matrix elements can be computed using the so-called Wick's
rule, which states that we must sum, over all possible pairings between $Z$'s and
$Z^\dag$'s, the product of the average values of the pairs. Namely,
\be\label{wick1}\left\llangle\prod_{k=1}^n
Z_{m_kj_k}Z^\dag_{q_{k}r_k}\right\rrangle=\sum_{\sigma\in S_n}\prod_{k=1}^n\llangle
Z_{m_kj_k}Z^\dag_{q_{\sigma(k)}r_{\sigma(k)}}\rrangle.\ee If we the quantity we wish to
average involves traces of $ZZ^\dag$, all we need to do is expand these traces in terms
of matrix elements and apply Wick's rule. Most importantly, we can then employ a
diagrammatic technique.

For example, suppose we wish to compute \be\label{avewick} \left\llangle {\rm
Tr}[(ZZ^\dag)^2]{\rm Tr}[(ZZ^\dag)^3]
Z_{i_1o_1}Z^\dag_{o_2,i_1}Z_{i_2o_2}Z^\dag_{o_1,i_2}\right\rrangle.\ee We start by
writing it as \be \sum_{m_1,...,m_5}\sum_{j_1,...,j_5}\left\llangle \left\{\prod_{k=1}^2
Z_{m_kj_k}Z^\dag_{j_k,m_{k+1}}\prod_{s=3}^5
Z_{m_sj_s}Z^\dag_{j_s,m_{s+1}}\right\}Z_{i_1o_1}Z^\dag_{o_2,i_1}Z_{i_2o_2}Z^\dag_{o_1,i_2}\right\rrangle,\ee
where all sums run from $1$ to $N$ (in the first product we mean $m_3\equiv m_1$, while
in the second product we mean $m_6\equiv m_3$). The diagrammatics consists in picturing
the matrix elements as pairs of arrows. Arrows that represent elements from $Z$ have a
marked end at the head, while arrows that represent elements from $Z^\dag$ have a marked
end at the tail. Arrows representing matrix elements coming from traces are arranged in
clockwise order around vertices, so that all marked ends are on the outside. Finally, the
elements that do not come from traces are arranged surrounding the other ones, also in
clockwise order. Since this is most easily explained by means of an image, we show it in
Figure 3(a).

Once we have arranged the arrows, Wick's rule consists in making all possible connections
between them, using the marked ends. Clearly, this produces a ribbon graph. According to
Eq.(\ref{cov}), when computing the value of a graph, each ribbon gives rise to a factor
$\Omega^{-1}$. For the example in Figure 3(a), there are 7! possible connections. We show
two of them in Figures 3(b,c). The coupling in Figure 3(b) leads to the identifications
\be i_1=m_1,\quad i_2=m_2=m_3=m_4=m_5 \quad o_1=j_2=j_3=j_4=j_5,\quad o_2=j_1,\ee and
gives a contribution of $\Omega^{-7}$ to the average (\ref{avewick}). Notice how this
coupling is similar to Figure 2. On the other hand, the coupling in Figure 3(c) leads to
the identifications \be i_1=m_1,\quad i_2=m_2=m_4=m_5 \quad o_1=j_2=j_3=j_5,\quad
o_2=j_1.\ee In this case the indices $m_3$ and $j_4$ remain free to be summed over.
Therefore, this coupling gives a contribution of $N^2\Omega^{-7}$ to the average
(\ref{avewick}).

\begin{figure}
\includegraphics[scale=0.7,clip]{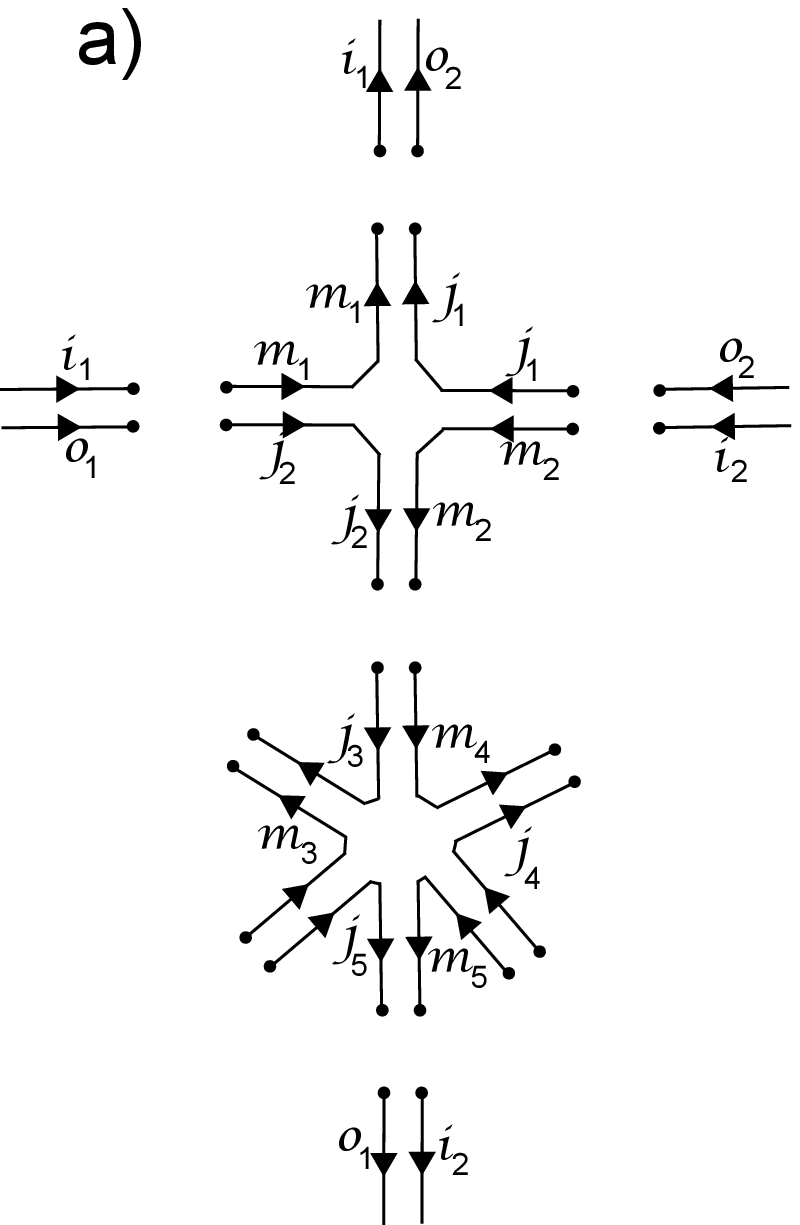}\quad \qquad\includegraphics[scale=0.55,clip]{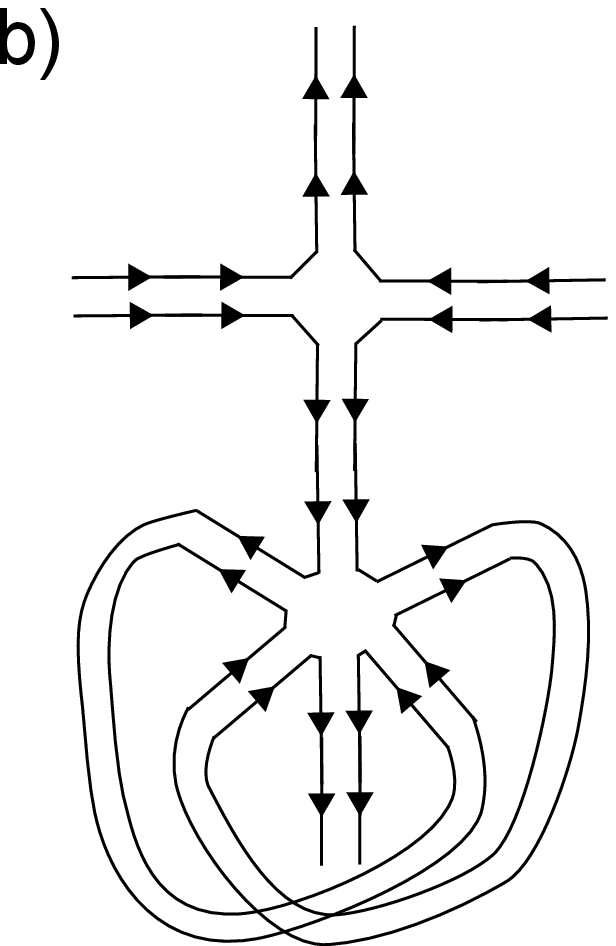}\quad \qquad\includegraphics[scale=0.55,clip]{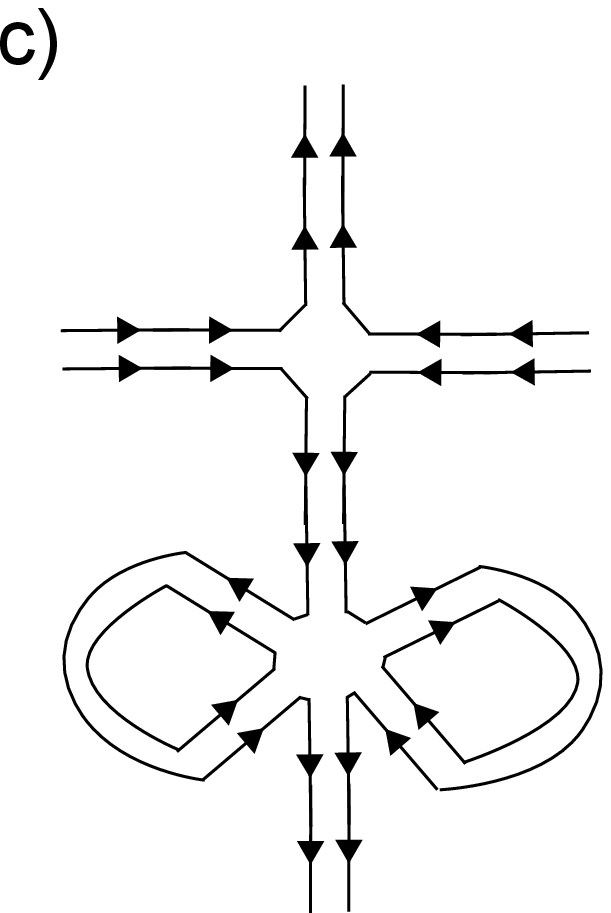}
\caption{Diagrammatics of Wick's rule, for the average in (\ref{avewick}). In a) we see
how the matrix elements are turned into arrows with marked ends and, in the case of
traces, arranged clockwise around vertices. In the vertex of valence 6 we have written
each label only once, for clarity. In b) and c) we see two particular Wick couplings, out
of the possible 7!. The labels of the arrows in b) and c) are the same as in a). Notice
the similarity between b) and Figure 2.}
\end{figure}

Free indices arise from closed loops in the ribbon graph. Each such loop increases by one
the number of faces of the graph (every graph has at least one face). Therefore, the
power of $N$ in the contribution of a given coupling is always one less than the number
of faces in the graph.

It should be clear that this theory is very close to the semiclassical approach to time
delay, provided we choose $\Omega=M(1-i\epsilon)$. However, the ribbon graphs in the
semiclassical theory always have a single face. As we have just mentioned, this
corresponds to keeping only those Wick couplings whose contribution does not depend on
$N$. Since all contributions are proportional to a positive power of $N$, we can simply
let $N\to 0$.

\subsection{Matrix integrals for correlation functions}

Let $\xi=(1\,2\,\cdots \, n)$ be the cyclic permutation of the first $n$ positive
integers,  and let $\vec{i}=(i_1,...,i_n)$ and $\vec{o}=(o_1,...,o_n)$. Introduce the
integral \be\label{ourintegral}
G_n(M,\epsilon,N,\vec{i},\vec{o})=\frac{1}{M\mathcal{Z}}\int dZ e^{-M\sum_{q\ge 1}
\frac{(1-iq\epsilon)}{q}\tr[(ZZ^\dag)^q]} \prod_{k=1}^n
Z_{i_ko_k}Z^\dag_{o_{\xi(k)}i_k}.\ee This can be seen as a Gaussian average as the ones
considered previously, if we understand the first term in the exponent,
$e^{-M(1-i\epsilon) \tr(ZZ^\dag)}$, to be part of the measure. Accordingly, we set
\be\label{norm} \mathcal{Z}=\int dZ e^{-M(1-i\epsilon)\tr(ZZ^\dag)}.\ee

The rest of the exponential can be Taylor expanded as \be e^{-M\sum_{q\ge 2}
\frac{(1-iq\epsilon)}{q}\tr[(ZZ^\dag)^q]}=\sum_{t=0}^\infty
\frac{(-M)^t}{t!}\left(\sum_{q\ge 2}
\frac{(1-iq\epsilon)}{q}\tr[(ZZ^\dag)^q]\right)^t.\ee For now, we consider this as a
formal power series and integrate term by term, employing Wick's rule and its
diagrammatical representation previously discussed. By construction, encounter vertices
of valence $2q$ will be accompanied by the factor $-M(1-iq\epsilon)$, giving the correct
semiclassical diagrammatic rules.

The integral (\ref{ourintegral}) is therefore designed to automatically produce all the
required ribbon graphs for the semiclassical evaluation of the correlation function
$C_n(\epsilon,M)$. The exponential produces all possible encounters, while the matrix
elements in the last product play the role of the channels. In line with
Eq.(\ref{multiple}), we must sum over all channels from $1$ to $M$, i.e. we must consider
the quantity \be\label{sumio}
\mathcal{G}_n(\epsilon,M,N)=\sum_{\vec{i},\vec{o}}G_n(\epsilon,M,N,\vec{i},\vec{o})\equiv
\sum_{i_1,\cdots,i_n=1}^{M}\sum_{o_1,\cdots,o_n=1}^{M}G_n(\epsilon,M,N,\vec{i},\vec{o}).\ee

The matrix integral produces more graphs than needed, but we have provided for this
overcounting. For example, the Taylor series of the exponential naturally has a $t!$ in
the denominator, which is responsible for eliminating the symmetry associated with
shuffling the vertices, when there are $t$ of them.  Also, graphs are produced that
differ from each other only by the rotation of a vertex. This is why we have divided
$\tr[(ZZ^\dag)^q]$ by $q$: it remedies the overcounting that would be caused by the
possible $q$ rotations of the vertex.

As we have discussed, in order to select only those ribbon graphs with a single face it
is necessary to take the limit $N\to 0$ at the end of the calculation. Therefore, the
correlation function will be given by \be C_n(\epsilon,M)=\lim_{N\to
0}\mathcal{G}_n(\epsilon,M,N).\ee

It is not very difficult to implement Eq.(\ref{ourintegral}) in a computer and obtain the
first few orders in $1/M$ for the first few correlation functions (the integral is not to
be performed numerically, of course, but using Wick's rule together with the covariance
(\ref{cov})). This leads to the results in (\ref{C1})-(\ref{C2}). Notice that letting
$N\to 0$ in this context presents no difficulty.

\section{Exact Solution}

This Section is dedicated to the exact solution of the matrix integral
(\ref{ourintegral}), and the calculation of its limit as $N\to 0$.

\subsection{Angular integration}

Introduce the singular value decomposition $Z=UDV$, where $D$ is real, positive and
diagonal while $U$ and $V$ are unitary. Let $X=D^2$ be a matrix with the same eigenvalues
as $ZZ^\dag$, and denote these eigenvalues by $x_i$, $1\leq i\leq N$. It is known
\cite{morris} that the measure $dZ$ is expressed in these new variables as \be
dZ=c_N|\Delta(X)|^2d\vec{x}dUdV,\ee where $c_N$ depends only on the dimension, $dU$ is
the normalized Haar measure on the unitary group $\mathcal{U}(N)$, and the Vandermonde
squared is the Jacobian of the transformation. This is a generalization of the
transformation from cartesian to polar coordinates in the complex plane. We shall first
perform the angular integration over $U$ and $V$.

A minor point to be mentioned is that $dV$ is not the same as the normalized Haar
measure. This is related to the fact that in the singular value decomposition there is a
certain ambiguity, as we may freely conjugate $D$ by a diagonal unitary matrix. The
matrix $V$ is thus uniquely determined only as an element of the coset
$\mathcal{U}(N)/[\mathcal{U}(1)]^N$. However, the functions we shall integrate,
polynomials in matrix elements as those in Section 3.3, are all invariant under
multiplication by a diagonal unitary matrix, and in this context $dV$ behaves just like
the Haar measure, up to normalization.

The only part of the integral in (\ref{ourintegral}) that depends on the angular
variables $U$ and $V$ is the last product. Thus, the angular integral we need is \be
\mathcal{A}=\int dUdV \prod_{k=1}^n\sum_{j_k,m_k}U_{i_kj_k}D_{j_k}V_{j_ko_k}
V^{\dag}_{o_{\xi(k)}m_k}D_{m_k}U^\dag_{m_k i_k}.\ee

Given $j=(j_1,j_2,...j_n)$, $m=(m_1,m_2,...,m_n)$ and $\tau\in S_n$, let \be
\delta_\tau[j,m]=\prod_{k=1}^n\delta_{j_km_{\tau(k)}}.\ee The Weingarten function of the
unitary group is defined by \be \int dU
\prod_{k=1}^nU_{a_kb_k}U^\dag_{c_kd_k}=\sum_{\sigma,\tau\in
S_n}\Wg_N(\tau\sigma^{-1})\prod_{k=1}^n \delta_\sigma[ad]\delta_\tau[bc],\ee and its
character expansion is \cite{samuel,esposti,collins} \be
\Wg_N(g)=\frac{1}{n!}\sum_{\substack{\lambda\vdash n\\\ell(\lambda)\leq N}}
\frac{d_\lambda}{[N]^\lambda}\chi_\lambda(g).\ee

Using the above machinery, we have \be \mathcal{A}=\sum_{\sigma\tau\rho\theta\in
S_{n}}{\rm Wg}^{U}_N(\rho\theta^{-1}){\rm
Wg}^{U}_N(\tau\sigma^{-1})p_{\tau^{-1}\theta}(X) \delta_\sigma[i, i]\delta_\rho[o,
\xi(o)],\ee where $p_\lambda$ are the power sum symmetric functions, and we have used
that \be \prod_{k=1}^n\sum_{j_k,m_k}D_{j_k}D_{m_k}\delta_\tau[j,m]\delta_\theta[j,m]=
\prod_{k=1}^n\sum_{j_k}x_{j_k}\delta_{\tau^{-1}\theta}[j,j]=p_{\tau^{-1}\theta}(X). \ee

The quantity we are after, Eq.(\ref{sumio}), requires summation over the indices
$\vec{i}$ and $\vec{o}$. It is easy to see that \be \sum_{i_1,\cdots,i_n=1}^{M}
\delta_\sigma[i, i]=M^{\ell(\sigma)}=p_{\sigma}(1^M), \ee where $\ell(\sigma)$ denotes
the number of cycles of the permutation $\sigma$, and \be \sum_{o_1,\cdots,o_n=1}^{M}
\delta_\rho[o, \xi(o)]=M^{\ell(\rho\xi)}=p_{\rho\xi}(1^M).\ee Notice that the channel
labels in the original matrix integral (\ref{ourintegral}) are all constrained to be
between $1$ and $N$. Nevertheless, we are summing them from $1$ to $M$. We are thus
assuming $N\geq M$. However, this will not deter us from letting $N\to 0$ later.

Once we expand \be p_{\tau^{-1}\theta}(X)=\sum_{\lambda\vdash
n}\chi_\lambda(\tau^{-1}\theta)s_\lambda(X),\ee where $s_\lambda$ are the Schur symmetric functions, we get \be
\sum_{\vec{i},\vec{o}}\mathcal{A}=\sum_{\lambda\vdash n}\sum_{\sigma\tau\rho\theta\in
S_{n}}{\rm Wg}^{U}_N(\rho\theta^{-1}){\rm
Wg}^{U}_N(\tau\sigma^{-1})\chi_\lambda(\tau^{-1}\theta)s_\lambda(X)p_{\sigma}(1^M)p_{\rho\xi}(1^M).\ee
Repeated use of the character orthogonality relation \be \sum_{\tau\in S_n}\chi_\mu(\tau) \chi_\lambda(\tau\sigma)=
\frac{n!}{d_\lambda}\chi_\lambda(\sigma)\delta_{\mu,\lambda}\ee leads to \be
 \sum_{\vec{i},\vec{o}}\mathcal{A}= \sum_{\lambda\vdash
n}\chi_\lambda(\xi)\left(\frac{[M]^\lambda}{[N]^\lambda}\right)^2s_\lambda(X).\ee

\subsection{Eigenvalue integration}

So far, the quantity we are after is given by \be\label{G}
\mathcal{G}_n(\epsilon,M,N)=\sum_{\vec{i},\vec{o}} G_n=\sum_{\lambda\vdash
n}\chi_\lambda(\xi)\left(\frac{[M]^\lambda}{[N]^\lambda}\right)^2\mathcal{R}(\epsilon,M,N),\ee where
$\mathcal{R}(\epsilon,M,N)$ is the radial integral over the eigenvalues of $ZZ^ \dag$. It is equal to
\be \mathcal{R}(\epsilon,M,N)=\frac{c_N}{M\mathcal{Z}}\int_0^1 d\vec{x}
\det{(1-X)^{M}}e^{Mi\epsilon{\rm Tr}\left[\frac{
X}{1-X}\right]}|\Delta(x)|^2s_{\lambda}(X),\ee where we have used that \be
e^{-M\sum_{q\ge 1}\frac{(1-iq\epsilon)}{q}\tr
X^q}=\det{\left[(1-X)^{M}\right]}e^{Mi\epsilon\tr\left(\frac{ X}{1-X}\right)}.\ee From
the well known Schur function expansion, \be e^{Mi\epsilon{\rm
Tr}\left(\frac{X}{1-X}\right)}=\sum_{m=0}^\infty \frac{(Mi\epsilon)^m}{m!}\sum_{\mu\vdash
m}d_\mu s_\mu\left(\frac{X}{1-X}\right),\ee we get
\be\label{radial}\mathcal{R}(\epsilon,M,N)=\sum_{m=0}^\infty \frac{(Mi\epsilon)^m}{m!}\sum_{\mu\vdash
m}d_\mu\mathcal{I}_{\lambda,\mu}(M,N),\ee where \be
\mathcal{I}_{\lambda,\mu}(M,N)=\frac{c_N}{\mathcal{Z}}\int_0^1 d\vec{x} \det
(1-X)^{M}|\Delta(x)|^2s_\mu\left(\frac{X}{1-X}\right)s_\lambda(X).\ee

Using the determinantal form of the Schur functions and the identity \be
\Delta\left(\frac{X}{1-X}\right)=\frac{\Delta(X)}{\det(1-X)^{N-1}},\ee one can show that \be \mathcal{I}_{\lambda,\mu}=
\frac{c_NN!}{\mathcal{Z}}\det\left(\frac{(M-\mu_j+j-1)!(\lambda_i-i+\mu_j-j+2N)!}{(M+2N+\lambda_i-i)!}\right).\ee
Two factorials can be taken out of the determinant, and we can write \be
\mathcal{I}_{\lambda,\mu}=\frac{c_NN!}{\mathcal{Z}}\prod_{j=1}^N
\frac{(M-\mu_j+j-1)!}{(M+2N+\lambda_j-j)!}\det\left((\lambda_i-i+\mu_j-j+2N)!\right).\ee
Introducing $(M+j-1)!$ in the product, we get
\be\mathcal{I}_{\lambda,\mu}=\frac{c_NN!}{\mathcal{Z}}\frac{1}{[M]_\mu}\prod_{j=1}^N
\frac{(M+N-j)!}{(M+2N+\lambda_j-j)!}\det\left((\lambda_i-i+\mu_j-j+2N)!\right).\ee

\subsection{The $N\to 0$ limit}

We must now take the $N\to 0$ limit. This is a delicate procedure. We can only do it for
quantities that are analytic functions of $N$. For example, using the singular value
decomposition, the normalization constant (\ref{norm}) becomes \be
\mathcal{Z}=c_N\int_0^\infty dxe^{-M(1-i\epsilon)\tr
X}|\Delta(x)|^2=\frac{c_N}{[M(1-i\epsilon)]^{N^2}}\prod_{j=1}^Nj!(N-j)!.\ee It is
perfectly fine to take the limit in the denominator. In the rest of the expression, we
must leave $N$ intact for now. In this sense, we write \be\label{norm2} \mathcal{Z}\to
c_N\prod_{j=1}^Nj!(N-j)!.\ee

The quantity $\mathcal{I}_{\lambda,\mu}$ contains the factor \be \prod_{j=1}^N
\frac{(M+N-j)!}{(M+2N+\lambda_j-j)!}.\ee First, we let $N\to 0$ inside the product, to
get\be \prod_{j=1}^N \frac{(M-j)!}{(M+\lambda_j-j)!}.\ee This still depends on $N$ via
the limit of the product. However, $\lambda_j=0$ for $j>\ell(\lambda)$. Hence, if we
assume $N\geq\ell(\lambda)$, we can write this as \be \prod_{j=1}^{\ell(\lambda)}
\frac{(M-j)!}{(M+\lambda_j-j)!}=\frac{1}{[M]^\lambda},\ee which is independent of $N$.
Now, in all rigor we are not allowed to take $N\to 0$ after assuming
$N\geq\ell(\lambda)$. We do it anyway, and write \be\mathcal{I}_{\lambda,\mu}\to
\frac{N!}{\mathcal{Z}}\frac{1}{[M]_\mu[M]^\lambda}\det\left((\lambda_i-i+\mu_j-j+2N)!\right).\ee

Further, we factor out the smallest factor from each row of the determinant, producing
$\prod_{j=1}^N(N+\lambda_j-j+\mu_N)!$. If we assume that $N>\ell(\mu)$, then $\mu_N=0$.
Hence, using (\ref{norm2}), \be\mathcal{I}_{\lambda,\mu}\to
\frac{N!}{[M]_\mu[M]^\lambda}\prod_{j=1}^N\frac{(N+\lambda_j-j)!}{(N-j)!j!}
\det\left(\frac{(\lambda_i-i+\mu_j-j+2N)!}{(\lambda_i-i+N)!}\right).\ee We again consider
$N\geq\ell(\lambda)$ first and $N\to 0$ later, to arrive
at\be\label{II}\mathcal{I}_{\lambda,\mu}\to
\frac{[N]^\lambda}{[M]_\mu[M]^\lambda}\frac{1}{\prod_{j=1}^{N-1}j!}\det\left(\frac{(\lambda_i-i+\mu_j-j+2N)!}{(\lambda_i-i+N)!}\right).\ee

\subsubsection{The determinant}

We need to consider the determinant
\be\mathcal{D}=\det\left(\frac{(\lambda_i-i+\mu_j-j+2N)!}{(\lambda_i-i+N)!}\right)
=\det\left(\frac{(a_i+b_j)!}{a_i!}\right),\ee where \be a_i=\lambda_i-i+N, \quad
b_j=\mu_j-j+N.\ee Each column consists of raising factorials, i.e. we have \be
(a_i+1)(a_i+2)\cdots(a_i+b_j)=\frac{[a_i]^{b_j+1}}{a_i}.\ee We therefore expand each
column using identity \be\label{Stir}
[x]^n=\sum_{k=0}^n\begin{bmatrix}n \\ k \end{bmatrix}x^k,\ee in terms of unsigned Stirling numbers of the first
kind. We get
\be \frac{[a_i]^{b_j+1}}{a_i}=\sum_{k_j=1}^{b_j+1}\left[\begin{array}{c}b_j+1 \\
k_j \end{array}\right]a_i^{k_j-1}=\sum_{k_j=0}^{b_j}\left[\begin{array}{c}b_j+1 \\
k_j+1 \end{array}\right]a_i^{k_j}.\ee The determinant is then given by
\be\mathcal{D}=\prod_{j=1}^N\sum_{k_j=0}^{b_j}\left[\begin{array}{c}b_j+1
\\k_j+1 \end{array}\right] \det\left(a_i^{k_j}\right).\ee Introducing $k_j=\omega_j-j+N$
we have \be\mathcal{D}=\prod_{j=1}^N\sum_{\omega_j=j-N}^{\mu_j}
\left[\begin{array}{c}\mu_j-j+N+1 \\ \omega_j-j+N+1 \end{array}\right]
\det\left(a_i^{\omega_j-j+N}\right).\ee

\begin{table}
\begin{tabular}{|c|l|ccccccccccc|}\hline
$F_{\lambda,\mu}$ &&&&&&&$\mu$ &&&
\\ \hline &$(1)$&1&1&1&2&1&2&6&2&0&2&6
\\ &$(11)$&2&4&2&12&4&4&48&12&0&8&12
\\ &$(2)$&2&2&4&4&4&12&12&8&0&12&48
\\ &$(111)$&3&9&3&36&9&6&180&36&0&18&18
\\ &$(21)$&3&6&6&18&12&18&72&36&0&36&72
\\ $\lambda$&$(3)$&3&3&9&6&9&36&18&18&0&36&180
\\ &$(1111)$&4&16&4&80&16&8&480&80&0&32&24
\\ &$(211)$&4&12&8&48&24&24&240&96&0&72&96
\\ &$(22)$&4&10&10&32&28&32&132&92&12&92&132
\\ &$(31)$&4&8&12&24&24&48&96&72&0&96&240
\\ &$(4)$&4&4&16&8&16&80&24&32&0&80&480 \\\hline
\end{tabular}
\caption{The function $F_{\lambda,\mu}$, for the first few values of $\lambda$ and $\mu$.}
\end{table}

Notice that $\omega$ is not a partition, since its elements are not necessarily ordered,
and they can be negative. Still, the last determinant, if it does not vanish, can be
turned into a Schur function by simply re-ordering the columns. Let $\widetilde{\omega}$
be the partition that is created in this way, and $|\widetilde{\omega}|$ the number it
partitions. For instance, if $\omega=(1,1,-1,1)$ we have \be \det\left(\begin{matrix}
a_i^N & a_i^{N-1} & a_i^{N-4} & a_i^{N-3} & a_i^{N-5} \cdots\end{matrix}\right)
=-\det\left(\begin{matrix} a_i^N & a_i^{N-1} & a_i^{N-3} \cdots\end{matrix}\right),\ee so
the corresponding partition is $\widetilde{\omega}=(1,1)$ and $|\widetilde{\omega}|=2$.
As we can see, the reordering of the columns may lead to a change in sign. Let
$\eta(\omega)$ denote this sign, so that \be
\det\left(a_i^{\omega_j-j+N}\right)=\eta(\omega)\Delta(a)s_{\widetilde{\omega}}(a)=\eta(\omega)\frac{d_\lambda}{n!}[N]^\lambda
s_{\widetilde{\omega}}(a)\prod_{j=1}^{N-1}j!.\ee

We must consider the $N\to 0$ limit of \be
s_{\widetilde{\omega}}(a)=\frac{1}{|\widetilde{\omega}|!}\sum_{\rho\vdash
|\widetilde{\omega}|}|\mathcal{C}_\rho|\chi_{\widetilde{\omega}}(\rho)p_\rho(a).\ee
 The
limit of $p_\rho(a)$ can be obtained simply removing from this quantity everything that
scales with $N$: \be\lim_{N\to
0}p_\rho(\{\lambda_i-i+N\})=\prod_{q=1}^{\ell(\rho)}\left(\sum_{i=1}^{\ell(\lambda)}
(\lambda_i-i)^q-(-i)^q\right)=:
f_\rho(\lambda).\ee We can finally write \be
\frac{\mathcal{D}}{\prod_{j=1}^{N-1}j!}\to\frac{d_\lambda}{n!}[N]^\lambda
F_{\lambda,\mu},\ee where the function $F_{\lambda,\mu}$ is given by \be
F_{\lambda,\mu}=\prod_{j=1}^{\ell(\mu)}\sum_{\omega_j=j-\ell(\mu)}^{\mu_j}
\left[\begin{array}{c}\mu_j-j+1 \\ \omega_j-j+1 \end{array}\right]
\frac{\eta(\omega)}{|\widetilde{\omega}|!}\sum_{\rho\vdash
|\widetilde{\omega}|}|\mathcal{C}_\rho|\chi_{\widetilde{\omega}}(\rho)f_\rho(\lambda).\ee

We tabulate some values of this function in Table 1. Several properties stand out from
inspection of these values. First, all values are non-negative integers. Second, and this
is easy to prove, that \be\label{mu1} F_{\lambda,(1)}=f_{(1)}(\lambda)=n \text{ if
}\lambda\vdash n.\ee Third, and this we can only conjecture, that \be
F_{\lambda,\mu}=F_{\lambda',\mu'},\ee where $\lambda'$ and $\mu'$ are the conjugate
partitions of $\lambda$ and $\mu$, respectively. Finally, we notice that, when $\mu$ is
not a hook partition, it seems that $F_{\lambda,\mu}$ is different from zero only if
$\lambda$ is also not a hook partition. We have checked this extensively, but do not have
a proof.

\subsection{Final Result}

It is time to put the pieces back together. We have to plug the limiting value of
$\mathcal{D}$ into the expression for $\mathcal{I}_{\lambda,\mu}$, Eq.(\ref{II}), put
this into the expression for the radial integral, Eq. (\ref{radial}), and finally arrive
at the quantity we want, which is $\mathcal{G}_n$, Eq.(\ref{G}). After some cancelations,
we get that the limit as $N\to 0$ of $\mathcal{G}_n$, which is nothing but the
semiclassical expression for the correlation function $C_n(\epsilon,M)$, is given by \be
\lim_{N\to 0}\mathcal{G}_n(\epsilon,M,N)=C_n(\epsilon,M)=\frac{1}{Mn!}\sum_{m=0}^\infty
\frac{(Mi\epsilon)^m}{m!}\sum_{\mu\vdash m}\sum_{\lambda\vdash n}d_\lambda
d_\mu\chi_\lambda(\xi)\frac{[M]^\lambda}{[M]_\mu} F_{\lambda,\mu}.\ee

This expression is perhaps not as simple we one might hope for, specially the
$F_{\lambda,\mu}$ part. This complication is probably due to the fact that we are using a
Taylor series in $\epsilon$. We know that, at each order in $1/M$, the correlation
functions are rational functions of $\epsilon$, with the denominator being a power of
$(1-i\epsilon)$. Maybe if this fact could be explicitly incorporated into the calculation
somehow, the resulting expression would be more manageable.

We can see from (\ref{mu1}) that $C_n(\epsilon,M)=1+ni\epsilon+O(\epsilon^2)$.

Using (\ref{MfromC}), our semiclassical approach leads to the following expression for the average
value of time delay moments $\mathcal{M}_m=\frac{1}{M}\tr (Q^m)$: \be \langle
\mathcal{M}_m\rangle=\frac{\tau_D^m M^{m-1}}{m!}\sum_{\mu\vdash m}\frac{d_\mu}{[M]_{\mu}}\sum_{n=1}^m
\frac{(-1)^{m-n}}{n!}{m \choose n}\sum_{\lambda\vdash n}d_\lambda\chi_\lambda(n)[M]^{\lambda}F_{\lambda,\mu}.\ee On the other hand, we have shown,\cite{previous} using the random
matrix theory approach, that \be \langle
\mathcal{M}_m\rangle_{\rm RMT}=\frac{\tau_D^m M^{m-1}}{m!}\sum_{\mu\vdash m}\frac{d_\mu}{[M]_{\mu}}\chi_\mu(m)[M]^{\mu}.\ee Therefore, if the identity
\be\label{identity} \sum_{n=1}^m \frac{(-1)^{m-n}}{n!}{m \choose n} \sum_{\lambda\vdash
n}d_\lambda\chi_\lambda(n)[M]^\lambda F_{\lambda,\mu}=\chi_\mu(m)[M]^\mu\ee holds,
the semiclassical formula for $\langle \mathcal{M}_m\rangle$ becomes exactly equal
to the corresponding RMT prediction. We have checked that (\ref{identity}) indeed holds for
all $\mu\vdash m$ up to $m=8$. This guarantees agreement between
the semiclassical and RMT calculations up to the first 8 moments. We expect that this
agreement should in fact hold for all moments, and in fact for all polynomial functions of the time delay matrix.

\section*{Acknowledgments}

Financial support from Conselho Nacional de Desenvolvimento Cient\'ifico e Tecnol\'ogico
(CNPq) is gratefully acknowledged.

\end{document}